\documentclass[aps,preprint]{revtex4}
\usepackage{graphicx,epsfig,subfigure,dcolumn,bm}

\begin{document}

\title{Microphase Separation Induced by Differential
Interactions in Diblock Copolymer/Homopolymer Blends}

\author{Jiajia Zhou}
\email[]{zhouj2@mcmaster.ca}
\author{An-Chang Shi}
\email[]{shi@mcmaster.ca}
\affiliation{Department of Physics and Astronomy, McMaster University \\
Hamilton, Ontario, Canada L8S 4M1}


\begin{abstract}
Phase behavior of diblock copolymer/homopolymer blends (AB/C) 
is investigated theoretically. 
The study focuses on a special case where all three binary 
pairs, A/B, B/C and C/A, are miscible.   
Despite the miscibility of the binary pairs, a closed-loop immiscible
region exists in the AB/C blends when the A/C and B/C pair interactions
are sufficiently different.   
Inside the closed-loop, the system undergoes microphase separation,
exhibiting different ordered structures. 
This phenomenon is enhanced when the homopolymer (C) interacts more
strongly to one of the blocks (A or B).
\end{abstract}


\maketitle

\section{Introduction}

The development of new polymeric materials is driven by the
increasingly complicated requirements of advanced engineering, 
as well as by the desire to improve material properties and
reduce production cost.
Beside synthesizing new types of homopolymers and copolymers, 
blending different polymers provides another route to obtain new
materials \cite{Paul_Newman, Utracki}.
Polymer blends can have combinative and enhanced properties of their 
components.
From a thermodynamic point of view, polymer blends may be
miscible, partially miscible, or immiscible.
The physical properties of polymer blends vary drastically 
in these different states. 
From this perspective, a good understanding of the phase behavior of
polymer blends is crucial. 
Because polymer blends are composed of more than one element, their
phase behavior is controlled by a large number of parameters, such as
chain lengths, monomer interactions and polymer architectures. 
Due to the very large parameter space, theoretical study is crucially
important to understand the phase behavior and material 
properties in this complex system.

The simplest polymer blend consists of two different homopolymers, A
and B.
For symmetric binary A/B blends, both homogeneous and inhomogeneous 
phases exist, depending on the interaction strength $\chi_{AB}N$, 
where $\chi_{AB}$ is the Flory-Huggins interaction parameter and 
$N$ is the degree of polymerization of the polymers \cite{deGennes}. 
For small values of $\chi_{AB}N$, the two homopolymers are miscible
and the blend is in a homogeneous state. 
For $\chi_{AB}N$ larger than a critical value ($\chi_{AB}N>2$), the
two homopolymers become immiscible. 
A macrophase separation occurs where the blends separate into A-rich 
and B-rich phases.
The situation is different when the A and B homopolymers are linked 
together to form AB diblock copolymers.
In this case, macrophase separation cannot take place because 
of the chemical connections between the A and B blocks. 
Instead, a microphase separation occurs in which the A and B are 
separated locally at a length scale determined by the size of the polymers
\cite{Leibler1980, Hamley}. 
The microphase separation is also controlled by the interaction
strength $\chi_{AB}N$.
For symmetric diblock copolymers, the critical value of $\chi_{AB}N$ 
for order-disorder transition is about 10.5.

It is interesting to blend AB diblock copolymers with homopolymers C. 
In this case the microphase separation of diblock copolymers can 
compete with the macrophase separation of the homopolymers.
The different interactions between different monomers provide a rich 
phase behavior.  
If the homopolymer C is immiscible with both blocks A and B, the 
situation is simple: the homopolymers will be separated from diblock 
copolymers. 
A more complex case involves AB/C blends where the homopolymers C are
immiscible with one block of the diblock copolymers, but interact
favorable with the other block \cite{Jiang1991, Jiang1995,
  Lowenhaupt1994, Zhao1997, Han2000}.  
This system is similar to the case of amphiphilic molecules in solution, 
which is important to understand the self-assembly of surfactants and
lipids. 
A simple example of this class of blends is that the homopolymers are
chemically identical to one of the blocks of the copolymers, i.e., an
AB/A blend.  
Binary AB/A blends have been extensively studied theoretically 
\cite{Semenov1993, Matsen1995, Matsen1995a, Janert1998}
and experimentally \cite{ Hashimoto1990, Tanaka1991, Winey1991}. 
The phase diagrams of AB/A blends show the coexistence of macrophase
separation and microphase separation, and the addition of the
homopolymers tends to stabilize some complex ordered structures.      

Another interesting case occurs for blends consisted of 
C homopolymers that are miscible with both the A and B blocks. 
The miscible nature of the blends may provide potential applications 
which rely on the homogeneity of the material.
Theoretically and practically, the extension to the attractive
interactions between compounds may lead to new phenomena 
and create new materials. 
In general, miscible blends are characterized by homogeneous phases. 
However, phase separation can be induced by differential
monomer-monomer interactions.  
An example is found in ternary blends composed of A/B/C 
homopolymers where all three binary pairs are miscible. 
A closed-loop immiscible region is found in the phase diagram 
\cite{Zeman1972, Jo1991, Manestrel1992, Kuo2002}. 
In cases that the attractive interaction between A/C is much stronger
than that of B/C and A/B, A and C homopolymers tend to be 
separated from B. 
On the other hand, macroscopic phase separation of A and B cannot take
place for AB/C blends, because A and B are chemically bonded together. 
Similar to the case of diblock copolymers, AB/C blends can reduce their
free energy through microphase separation.  
Indeed, the microphase separation of AB/C blends has been observed in
experiment of Chen et al. \cite{Chen2008}. 
These authors investigated AB/C blends where the interactions 
between each pair of segments are favorable. 
Their experiment revealed a phase diagram with a closed microphase 
separation loop, despite the fact that the C homopolymers have attractive
interactions with both blocks of the copolymers.

In this paper, we present a theoretic study of the phase behavior of 
diblock copolymer/homopolymer blends.
The study focuses on the case where the homopolymers attract to 
one block much strongly than to the other block of the diblock copolymers. 
We use the Gaussian chain model for the polymers, and
apply the random phase approximation (RPA) 
to examine the stability limits of the homogeneous phase, leading to
both macrophase and microphase separation transition.
While the RPA calculates the stability boundary of the homogeneous 
phase, we are also interested in the morphological details inside 
the microphase separation region. 
For this purpose, self-consistent field theory is employed.
Various representative phase diagrams are constructed to
illuminate the general trends in the phase behavior by varying
the homopolymer length and monomer-monomer interaction parameters.
Comparison with available experiments is given.

\section{RPA Analysis}

The system discussed in this work is a mixture of AB diblock
copolymers and C homopolymers. 
For the diblock copolymer, the degrees of polymerization for the A- 
and B-block are $Nf_A$ and $N(1-f_A)$, respectively, and the C
homopolymers are with a degree of polymerization $\kappa N$. 
The volume fraction of the diblock copolymers and homopolymers 
in the blends are $1-\phi_H$ and $\phi_H$, respectively. 
For simplicity we assume all species have the same monomer volume 
$\rho_0^{-1}$ and statistical segment length $b$. 
The interactions between each pair of monomers are characterized 
by three Flory-Huggins parameters, $\chi_{AB}$, $\chi_{BC}$ and $\chi_{AC}$.

The RPA analysis of the system \cite{deGennes, Leibler1980, Mori1987,
  Ijichi1988} starts with an external potential $u_i$ acting on the
$i$ monomer.  
Assuming that the external fields are small, the density response
$\delta \phi_i$ to the external potential  can be written in the 
Fourier space as a linear function of the external potential, 
\begin{equation}
  \label{eq:phi_tildeS}
  \delta \phi_i (q) = - \beta \sum_{j} \tilde{S}_{ij}(q) u_j(q),
\end{equation}
where $\tilde{S}_{ij}(q)$ is the Fourier transform of the
density-density correlation between $i$ and $j$ monomers. 
Here we have used the fact that functions depend on the wave vector
$\mathbf{q}$ only through its magnitude $q \equiv |\mathbf{q}|$ because of
the isotropic symmetry of a homogeneous phase.

The interaction between the monomers can be taken into account 
through the mean-field approximation. 
To this end, the effective potential acting on the $i$ monomer is
written in the form, 
\begin{equation}
  \label{eq:ueff}
  u^{\mathrm{eff}}_i (q) = u_i (q) + \beta^{-1} \sum_{j \ne i} \chi_{ij} \delta
  \phi_j + \gamma,
\end{equation}
where the second term accounts for the mean-field interaction between
$i$ and $j$ monomers, and $\gamma$ is a potential required to assure
the incompressible condition 
\begin{equation}
  \label{eq:incompressible}
  \sum_i \delta \phi_i (q) = 0.
\end{equation}

The random phase approximation assumes that the density response
$\delta \phi_i$ is given by the effective potential, 
\begin{equation}
  \label{eq:phi_S}
  \delta \phi_i (q) = - \beta \sum_{j} S_{ij}(q) u_j^{\mathrm{eff}}(q),
\end{equation}
where $S_{ij}(q)$ is the Fourier transform of the density-density
correlation when the external potential vanishes.

Equations (\ref{eq:phi_tildeS})-(\ref{eq:phi_S}) form a set of
simultaneous equations for the unknowns $\delta \phi_i$ and
$\gamma$. 
By solving these equations, the correlation functions $\tilde{S}_{ij}$ 
can be found in terms of $S_{ij}$ and $\chi_{ij}$. 
The final results are lengthy and details can be found 
in the work of Ijichi and Hashimoto\cite{Ijichi1988}. 

The free energy of AB/C blends can be written as a Landau expansion
about the homogeneous state in terms of the density fluctuations
$\delta\phi_i$,
\begin{eqnarray}
\label{eq:2nd_fE}
\frac{F- F_{\mathrm{hom}}}{(\rho_0 V / \beta N)} && = \frac{1}{2! (2\pi)}
\int \tilde{S}^{-1}_{ij} (q) \delta\phi_i(q) \delta\phi_j(-q)
\mathrm{d} q + \cdots,  \nonumber \\ 
&& = \frac{1}{2! (2\pi)} \int \lambda_k (q) |\delta\psi_k(q)|^2
\mathrm{d}q + \cdots,  
\end{eqnarray}
where $\tilde{S}^{-1}_{ij} (q)$ is the inverse of the correlation
function $\tilde{S}_{ij} (q)$. 
Here we have neglected terms with order higher than two, and the
second-order term can be further written in a quadratic form, where 
$\lambda_k(q)$ are eigenvalues of matrix $\tilde{S}^{-1}_{ij}(q)$. 
The stability of the homogeneous phase depends on the sign of
eigenvalues $\lambda_k(q)$.  
When $\lambda_k(q)>0$, the contribution of any fluctuations to the
free energy is always positive, so the homogeneous phase is stable.
When $\lambda_k(q)<0$, the fluctuations reduce the free energy
and the homogeneous phase is unstable.

\begin{figure}[htp]
  \centering
  \includegraphics[width=0.9\columnwidth]{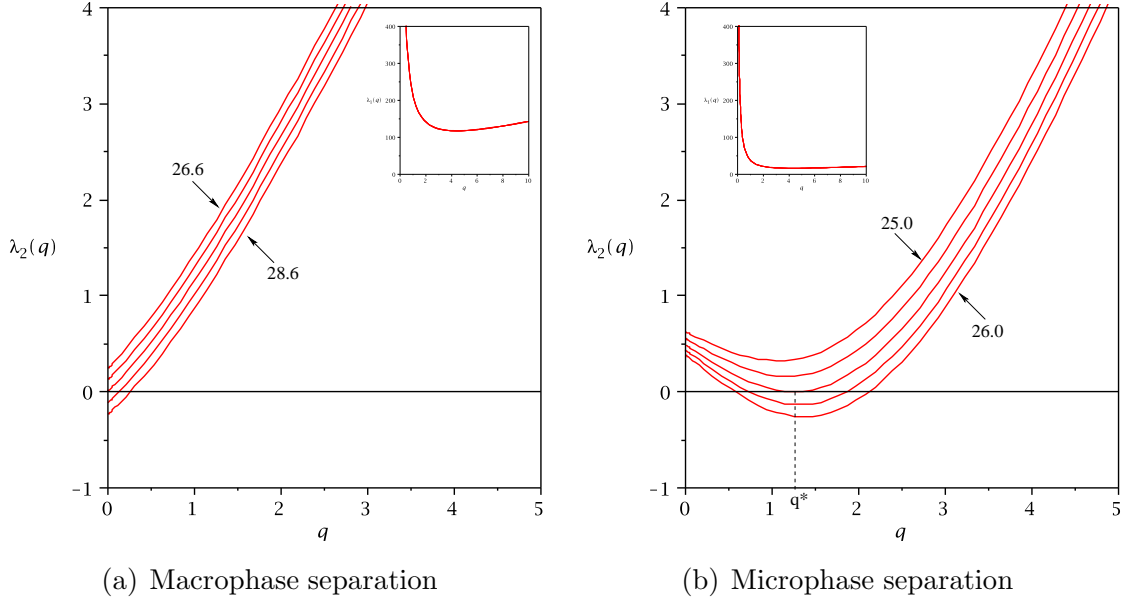}
  \caption{
    Plots of eigenvalues $\lambda_k(q)$ for (a) $\kappa=1.0$,
    $f_A=0.2$, $\phi_H=0.8$, $\chi_{AB}N=15$, $\chi_{BC}N=0$ and a
    range of $\chi_{AC}N$ values (26.6 $\sim$ 28.6), and (b) $\kappa=1.0$,
    $f_A=0.2$, $\phi_H=0.2$, $\chi_{AB}N=15$, $\chi_{BC}N=0$ and a
    range of $\chi_{AC}N$ values (25.0 $\sim$ 26.0). The insets show
    the positive eigenvalue $\lambda_1(q)$. Variations of
    $\lambda_2(q)$ with $\chi_{AC}N$ are shown, but they are not
    visible for $\lambda_1(q)$. The unit of the wave vector $q$ is
    $R^{-1}_g$, where $R_g=\sqrt{Nb^2/6}$ being the radius of
    gyration.
  }
  \label{fig:lambda}
\end{figure}

Normally, the inverse of the correlation function $\tilde{S}^{-1}_{ij}(q)$ 
is a 3$\times$3 matrix for the AB/C blends.
The incompressible condition reduces the order of the matrix by one. 
The spinodal line is determined by the condition that the smaller 
eigenvalue goes to zero. 
Typical plots of $\lambda_k(q)$ are shown in Figure \ref{fig:lambda}. 
In general, one of the eigenvalues $\lambda_1(q)$ is always positive,
while the other one $\lambda_2(q)$ approaches zero when $\chi_{AC}N$
changes. 

The macrophase separation is characterized by $\lambda_2(q)
\rightarrow 0$ at $q=0$. 
This is shown in Figure \ref{fig:lambda}(a) for blends with
$\kappa=1.0$, $f_A=0.8$, $\phi_H=0.2$, $\chi_{AB}N=15$, 
$\chi_{BC}N=0$ and different $\chi_{AC}N$ values.
The eigenvalue $\lambda_2(q)$ has a minimum at $q=0$, and the minimum
value approaches zero when $\chi_{AC}N$ increases.  
When $\lambda_2(q)$ becomes negative, any small density fluctuations
with $q=0$ decrease the free energy, leading to a growth of
fluctuations with macroscopic wavelength.
This corresponds to a macrophase separation between the diblock-rich
phase and homopolymer-rich phase.

On the other hand, the diblock copolymers in the blend introduce 
the possibility of microphase separation.
A microphase transition is characterised by the eigenvalue
$\lambda_2(q) \rightarrow 0$ at some finite $q^*>0$.  
This is shown in Figure \ref{fig:lambda}(b) for 
blends with $\kappa=1.0$, $f_A=0.2$, $\phi_H=0.2$, $\chi_{AB}N=15$,
$\chi_{BC}N=0$ and different $\chi_{AC}N$ values. 
The Fourier mode with nonzero wave number $q^*$ becomes unstable upon
increasing $\chi_{AC}N$, leading to the formation of ordered structure 
with length scale of $(q^*/2\pi)^{-1}$. 
This is in contrast to Figure \ref{fig:lambda}(a) where the Fourier
mode $q=0$ is destabilized first.

The six parameters $(\kappa, f_A, \phi_H, \chi_{AB}N, \chi_{BC}N, 
\chi_{AC}N)$ characterising the AB/C blends lead to a huge phase space. 
Some restrictions are needed so that the phase behavior can be described. 
We will keep $\kappa=1.0$ for the RPA calculation, which means the
homopolymer C has the same degree of polymerization as the diblock
copolymer AB.  
Furthermore, we will assume $\chi_{BC}N=0$, which represents the case
where the A/C interaction is much stronger than the B/C interaction.  

\begin{figure}[htp]
  \centering
  \includegraphics[width=0.9\columnwidth]{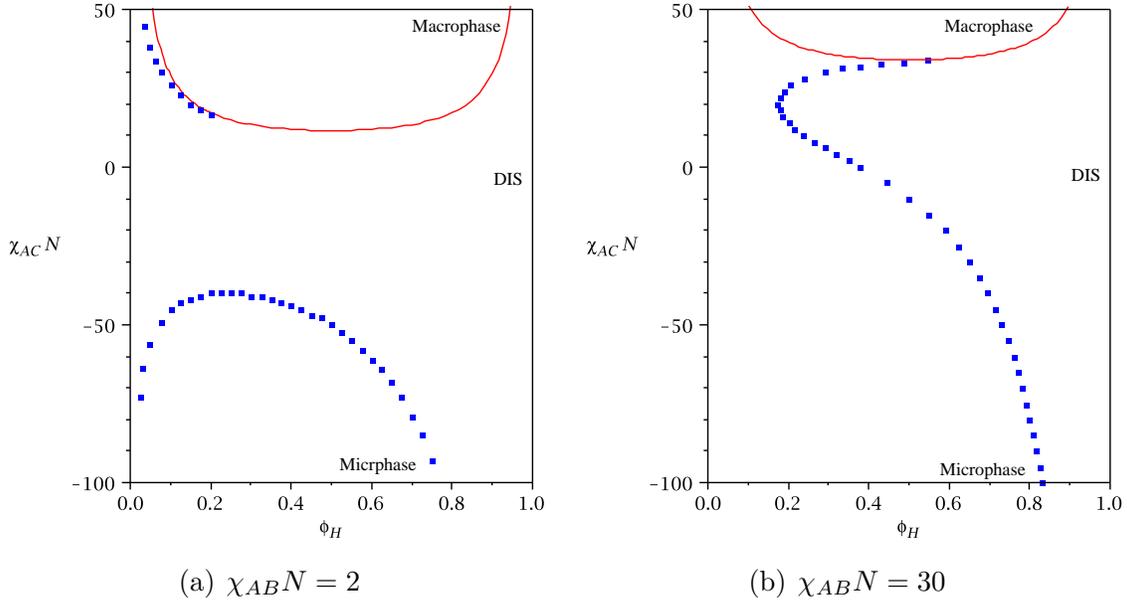}
  \caption{
    Phase diagrams for blends with parameters $\kappa=1.0$,
    $f_A=0.2$, $\chi_{BC}N=0$ and (a) $\chi_{AB}N=2$, (b)
    $\chi_{AB}N=30$.  The solid and dotted lines represent the
    stability limits for macrophase and microphase separation
    transition, respectively. Regions of macrophase separation,
    microphase separation and disordered states are labeled by
    Macrophase, Microphase and DIS. 
  }
  \label{fig:xABN}
\end{figure}

Figure \ref{fig:xABN}(a) shows a typical phase diagram in the 
$\phi_H$-$\chi_{AC}N$ plane. The parameters are $f_A=0.2$ and
$\chi_{AB}N=2$. 
The solid lines and dotted lines represent, respectively, the
stability limits for the macrophase separation transition
$(\chi_{AC}N)_{\mathrm{macro}}$ and microphase separation transition
$(\chi_{AC}N)_{\mathrm{micro}}$. 

The blends are disordered around $\chi_{AC}N=0$ and $\phi_H=1$, 
and ordered phases appear at both $\chi_{AC}N>0$ 
and $\chi_{AC}N<0$. 
In the region where $\chi_{AC}N>0$, increasing $\chi_{AC}N$ 
induces an instability to either macrophase separation or microphase 
separation, depending on the blend composition and 
copolymer asymmetry. 
For blends with a minority of homopolymers, microphase separation
occurs first when $\chi_{AC}N$ is increased. 
However, when the composition of the homopolymers increases to certain
value, the blend undergoes macrophase separation when $\chi_{AC}N$ is
increased. 
The critical composition for blends with $f_A=0.2$ 
is $\phi_H \approx 0.2$. 
The shape of the stability lines also suggests that, for 
small $\phi_H$ value, both $(\chi_{AC}N)_{\mathrm{macro}}$ 
and $(\chi_{AC}N)_{\mathrm{micro}}$ decrease with increasing 
$\phi_H$, while $(\chi_{AC}N)_{\mathrm{macro}}$ increases 
with increasing $\phi_H$ at large $\phi_H$ value. 
The former case suggests that the composition of the 
diblock copolymer in the blend 
suppresses the phase transition, which is known as the 
compatibilizing effect of the block polymers \cite{Tanaka1988}. 

On the other hand, the microphase separation also occurs 
in the region where $\chi_{AC}N$ is negative.
Since the interaction between the blocks $\chi_{AB}N=2$, the pure
diblock copolymers are in the disordered state.
The stability line for the microphase separation approaches $\phi_H=0$
axis infinitely close when $\chi_{AC}N$ has a large negative value.  
It is interesting that for a homogeneous diblock copolymer melt, 
adding a small amount of homopolymers C which has a strong
attractive interaction with one of the blocks will induce the phase
separation.  
Upon increasing $\phi_H$, $(\chi_{AC}N)_{\mathrm{micro}}$ increases 
sharply at first, then decreases after $\phi_H \approx 0.25$. 

A similar phase diagram for blends with $\chi_{AB}N=30$ is shown in
Figure \ref{fig:xABN}(b).  
In this case, the pure diblock copolymers are in the ordered state. 
This leads to the convergence of the two microphase separation regions
at $\chi_{AC}N>0$ and $\chi_{AC}N<0$. 

\begin{figure}[htp]
  \centering
  \includegraphics[width=0.9\columnwidth]{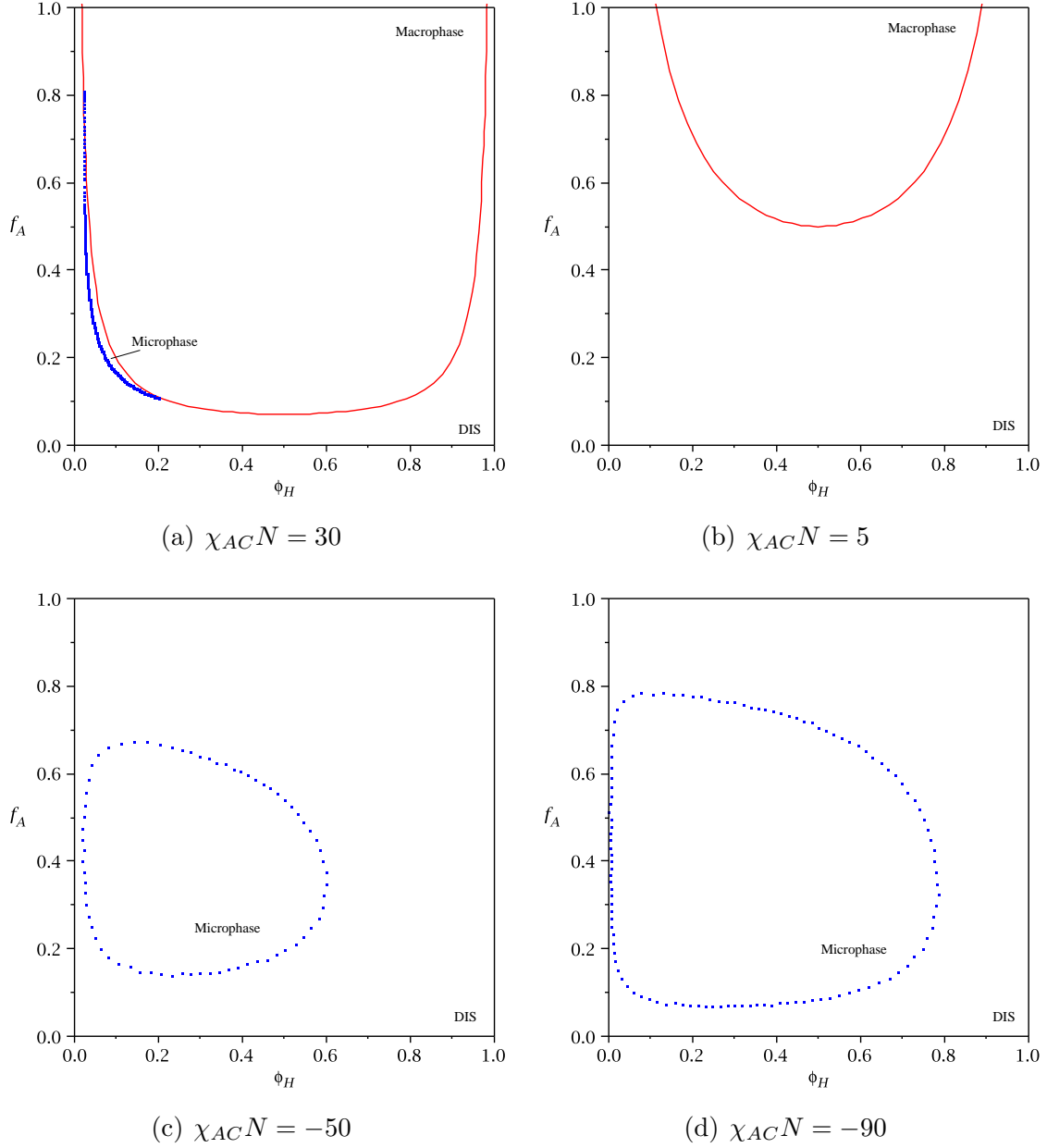}
  \caption{
    Phase diagrams for AB/C blends in $\phi_H$-$f_A$ plane. 
    The parameters are $\kappa=1.0$, $\chi_{AB}N=2$ and
    $\chi_{BC}N=0$. The solid and dotted lines represent the
    transitions for macrophase and microphase separation,
    respectively.
  }
  \label{fig:xACN}
\end{figure}

Another perspective to understand the phase transition is 
to plot the phase diagram in the $\phi_H$-$f_A$ plane 
for fixed interaction parameters.  
Figure \ref{fig:xACN} shows phase diagrams for blends with
$\chi_{AB}N=2$ and different values of $\chi_{AC}N$. 
They can be viewed as cross-sections pictures of 
Figure \ref{fig:xABN}(a) at different values of $\chi_{AC}N$. 
At $\chi_{AC}N=30$, because of the strong
repelling interaction between A/C, large region of macrophase
separation exists, while a small region of microphase separation
occurs near the $\phi_H=0$ axis.   
As $\chi_{AC}N$ decreases, the macrophase separation region shrinks 
and eventually disappears at negative value of $\chi_{AC}N$. 
At the same time, a closed-loop microphase separation region
appears, as can be seen from Figure \ref{fig:xACN}(c) and
\ref{fig:xACN}(d).  
Qualitatively, this closed immiscible loop corresponds to the one
observed in experiments \cite{Chen2008}.    

\section{Self-Consistent Field Theory}

Because of its simplicity, random phase approximation provides a
convenient method to calculate the order-disorder transition for
various parameter sets, but it is difficult to apply RPA to the
order-order transition.  
On the other hand, self-consistent field theory has been proven a
powerful method to determine the microstructures of polymer blends
\cite{Matsen1995, Matsen1995a, Janert1998, Matsen1994,
  Shi2004_chapter, Fredrickson}.   
Therefore, it is desirable to apply self-consistent field theory to
the study of the phase behavior of AB/C blends.

In the framework of self-consistent field theory, the system that 
consists of many interacting diblock copolymer/homopolymer 
chains is replaced by the problem of an ideal Gaussian chain 
in an averaged effective mean-field potential which depends on
the position of the chain. 
Once the mean-field potential is specified, the thermodynamics
properties of the system, such as the partition function and monomer 
densities, can be expressed in terms of the chain propagators, 
which are related to the potential by a modified diffusion equations.
From the monomer densities, a new mean-field potential can be
constructed. 
The procedure results a closed set of equations that can 
be solved self-consistently using numerical methods.
Once the self-consistent solutions are reached, the free energy 
can be estimated for various ordered structures.
A phase diagram is constructed by finding the morphology with the
lowest free energy. 
In this work, the calculation is performed in the grand-canonical
ensemble, and the self-consistent equations are solved using 
the spectral method \cite{Matsen1994,Laradji1997, Shi2004_chapter}.

For simplicity, we consider only the classical phases in the current
study: lamellae (LAM), cylinders on a hexagonal lattice (HEX), 
and spheres on a body-centered cubic lattice (BCC). 
Complex structures, such as close-packed spheres, 
perforated lamellae and bi-continuous cubic phases, 
can also occur for certain parameters \cite{Matsen1995}. 
These non-classical phases were found only in narrow regions 
between the classical phases.  
In this study, we are more interested in the evolution of 
the phase diagram by varying different parameters, 
therefore we restrict ourselves to the three classical phases. 

\begin{figure}[htp]
  \centering
  \includegraphics[width=0.65\columnwidth]{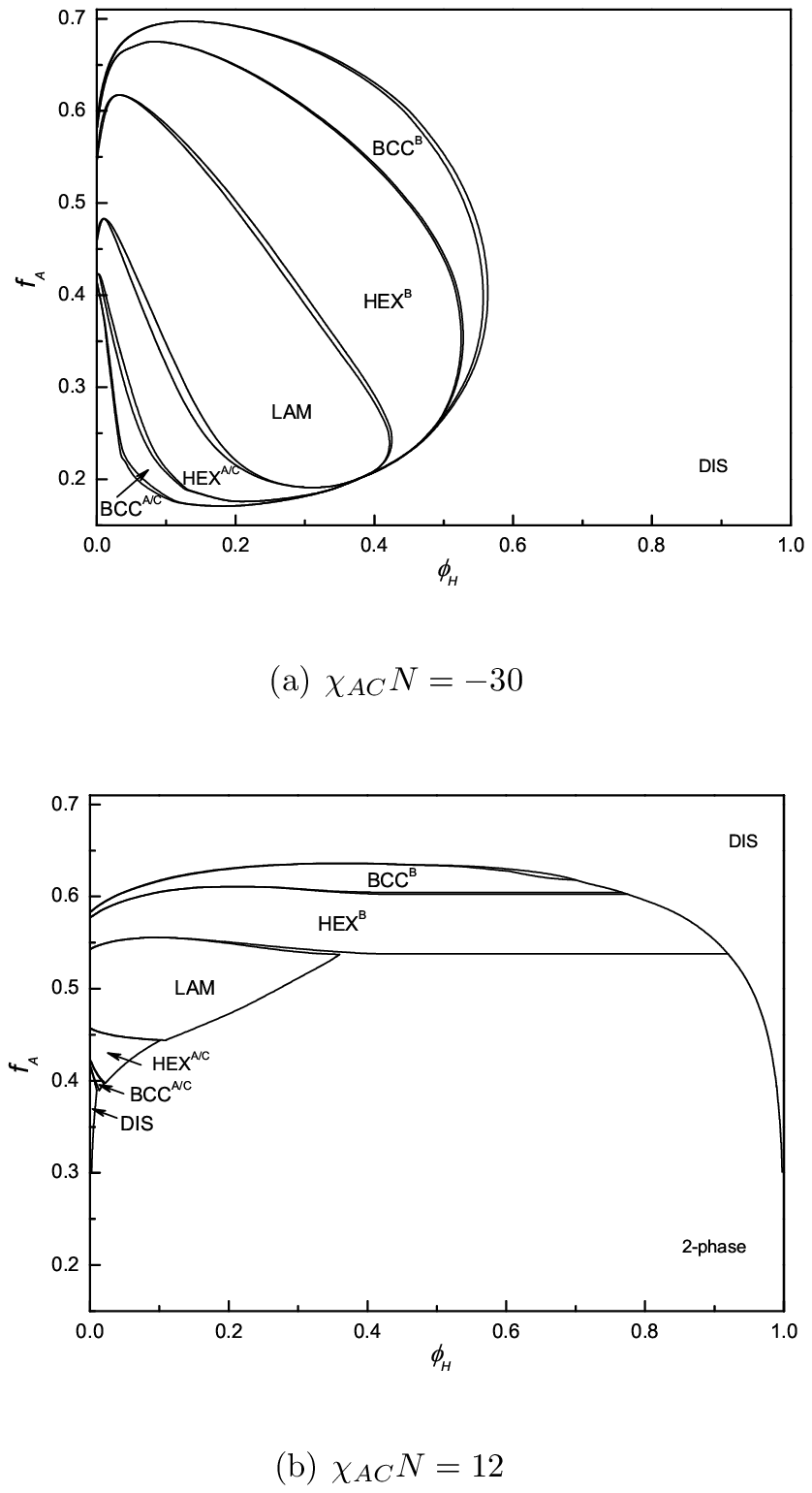}
  \caption{
    Phase diagrams of AB/C blends with parameters $\kappa=1.0$,
    $\chi_{AB}N=11$, $\chi_{BC}N=0$, and (a) $\chi_{AC}N=-30$, (b) 
    $\chi_{AC}N=12$. The superscripts of HEX and BCC phase denote the
    components that form the cylinders and spheres near the lattice
    centers. There are regions of 2-phase coexistence between the
    ordered phases. 
  }
  \label{fig4}
\end{figure}

Figure \ref{fig4}(a) shows a phase diagram in the $\phi_H$-$f_A$ plane 
for blends with $\kappa=1.0$, $\chi_{AB}N=11$, $\chi_{BC}N=0$ and
$\chi_{AC}N=-30$. 
In this case, the value of $\chi_{AB}N$ is large enough so the diblock
copolymer melt is in an ordered state.  
The results should converge to the pure diblock results as 
the homopolymer concentration goes to zero, which is the case as 
shown in Figure \ref{fig4}(a).
The phase behavior shown in Figure \ref{fig4}(a) is typical for the
case where the homopolymers attract strongly to one of the blocks
($\chi_{AC}N=-30$).  
This attractive interaction drives the C homopolymers to the
A-domains, leading to larger regions of ordered phases for mediated
homopolymer concentrations. 
For comparison, a phase diagram for the blends with $\kappa=1.0$,
$\chi_{AB}N=11$, $\chi_{BC}N=12$ and $\chi_{AC}N=0$ is presented in
Figure \ref{fig4}(b).  
Here the parameters are chosen in such a way that the C monomers 
accumulate inside the A-rich region for both cases.
Figure \ref{fig4}(b) is similar to the results of AB/A blends
from Matsen \cite{Matsen1995a} because the interaction parameters
$\chi_{AB}N$ and $\chi_{BC}N$ are close to each other, thus the system
resembles the AB/A blends. 

These two phase diagrams display similar features when the homopolymer
concentration is small. 
In this region, the majority of the blend is diblock copolymers and
adding small amount of homopolymers should preserve the morphologies. 
When the homopolymer concentration becomes large, these two phase
diagrams exhibit different phase behavior.  
For the case shown in Figure \ref{fig4}(b), where the interaction
parameter $\chi_{BC}N$ is positive, two possible scenarios of phase
behavior appear.   
When $f_A$ is large, only one disordered phase is stable. 
The blend tends to undergo unbinding transition where the spacing of
the ordered structure increases and eventually diverges at certain
homopolymer concentration \cite{Matsen1995, Matsen1995a, Janert1998}. 
When $f_A$ is small, macroscopic phase separation becomes an option. 
Adding more homopolymers induces a macrophase separation transition,
in which the blend separates into copolymer-rich and homopolymer-rich
regions.  
On the other hand, for the case shown in Figure \ref{fig4}(a), where
$\chi_{AC}N$ has a large negative value, only one disordered phase is
permitted.  
When the homopolymer concentration is large, the blends exhibit neither 
the unbinding transition nor the macrophase separation. 
This behavior is due to the strong attractive interaction between the
A and C monomers.   
As the homopolymer concentration is increased close to unity, both
cases reach a disordered phase.   

\begin{figure}[htp]
  \includegraphics[width=0.65\columnwidth]{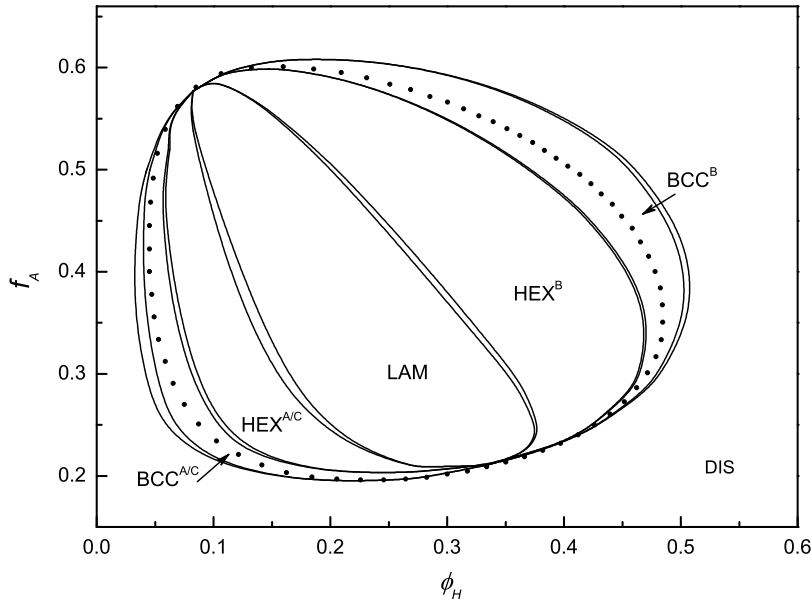}
  \caption{
    Phase diagram of AB/C blends with parameters
    $\kappa=1.0$, $\chi_{AB}N=2$, $\chi_{BC}N=0$ and
    $\chi_{AC}N=-40$. The dotted line shows the RPA result.
  } 
  \label{fig5}
\end{figure}

When all three binary pairs are miscible, the AB/C blends exhibit
another type of phase behavior.  
Figure \ref{fig5} shows a typical phase diagram for this case with
$\kappa=1.0$, $\chi_{AB}N=2$, $\chi_{BC}N=0$ and $\chi_{AC}N=-40$. 
In this case the blends are in a disordered phase at $\phi_H=0$ and
$\phi_H=1$.  
Ordered phases occur in a closed-loop region, as indicated by the RPA
analysis.  
Inside the closed-loop, different ordered structures are found. 
The order-order transitions between these structures are controlled
largely by the homopolymer concentration $\phi_H$.  
 
There are two critical points at which all ordered phases
converge on the order-disorder transition boundary.  
For the parameters used here, one point is located at
($\phi_H=0.083$, $f_A=0.58$), and the other one at ($\phi_H=0.33$,
$f_A=0.21$). 
The order-disorder transition is a second-order transition at these two
critical points, while the transition is first-order elsewhere.   
The nature of the order-disorder transition (ODT) can be understood by
considering the third-order term in the free energy expansion.  
A first-order phase transition occurs when a non-zero third-order term
is present. 
In the case of a phase transition from the disordered phase to the BCC
phase, the third-order term does not vanish in general.  
Therefor a first-order ODT is expected.
At the critical points, the ODT corresponds to a direct transition
from the disordered phase to a lamellar phase. 
In this case the symmetry of the lamellar phase ensures the vanishment
of the third-order term, resulting in a second-order ODT
\cite{Wang2009}.

It is important to point out the limitation of the present theory. 
The mean-field approach used here approximates the fluctuating
potential by a thermal averaged potential.  
This assumption becomes inaccurate near the order-disorder transition,
where the fluctuations are large.  
The fluctuations can shift the position of the phase boundary, 
and may change the second-order ODT into a weakly first-order transition
\cite{Fredrickson1987, Kielhorn1997}.

When the homopolymer concentration is increased, the AB/C blends
change from a disordered phase to BCC, HEX and LAM phases.  
This sequence of phase transition can be understood by examining 
how homopolymers are distributed in the diblock melt.  
Since the interaction between A and B are at $\chi_{AB}N=2$, which is
well below the ODT value of 10.5 for symmetric diblocks, the pure
diblock copolymers are in a disordered state.    
When small amount of homopolymers are added to the system, due to
their strong attraction to A-blocks, the homopolymers are distributed
around the A-blocks, resulting in an effective segregation of the A-B
blocks.  
Ordered phases emerge when this effective segregation becomes strong
enough.  
Another way to picture this effect is to take the homopolymers as the
core-forming agents of micelles in a diblock copolymer melt.  
The appearance of the spherical phase as the homopolymers are added
can be regarded as the ordering of these micelles. 
The key observation here is that the reduction of interaction energy
from A-C segregation is sufficient to overcome the entropy loss.  
At the strong-segregation limit, a similar mechanism for the formation
of reverse structures has been discussed by Semenov
\cite{Semenov1993}.  

At higher homopolymer concentration, the volume of A/C domain increases. 
The packing requirement of larger A/C domains leads to a change of the
interfaces between A/C and B, such that the interfaces are curved more
towards the B-domain.  
This mechanism lead to order-order phase transitions from BCC to HEX,
and then from HEX to LAM phases.   
After the LAM structure, adding more homopolymers induces the phase
transition to HEX, and then to BCC, where the HEX
and BCC both have B-blocks as the central components.
Eventually, a disordered phase is reached when sufficient amount of
homopolymers are added. 

\begin{figure}[htp]
  \includegraphics[width=0.65\columnwidth]{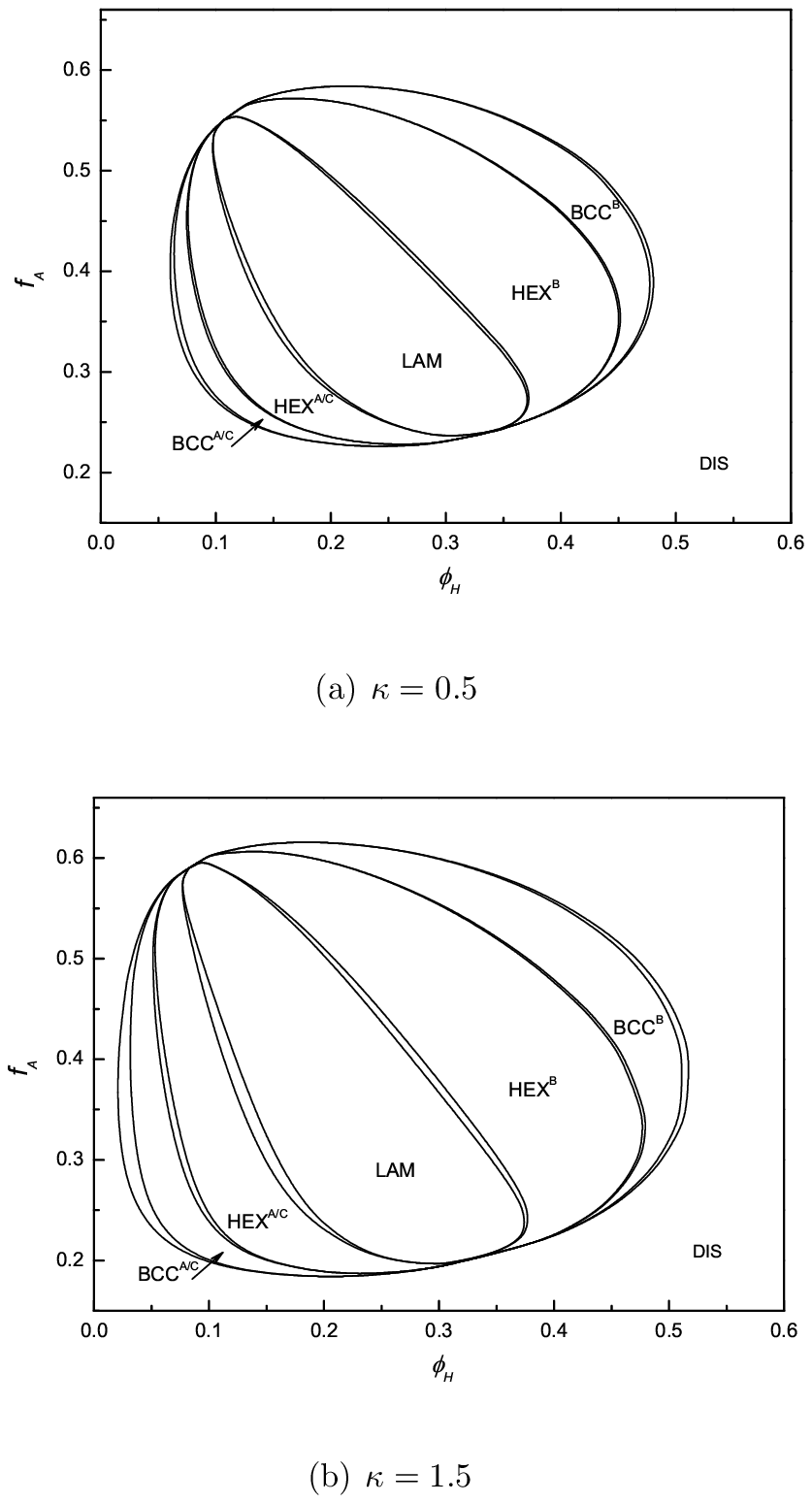}
  \caption{
    Phase diagrams of AB/C blends with parameters
    $\chi_{AB}N=2$, $\chi_{BC}N=0$, $\chi_{AC}N=-40$ and
    (a) $\kappa=0.5$, (b) $\kappa=1.5$.
  }
  \label{fig6}
\end{figure}

\begin{figure}[htp]
  \includegraphics[width=0.65\columnwidth]{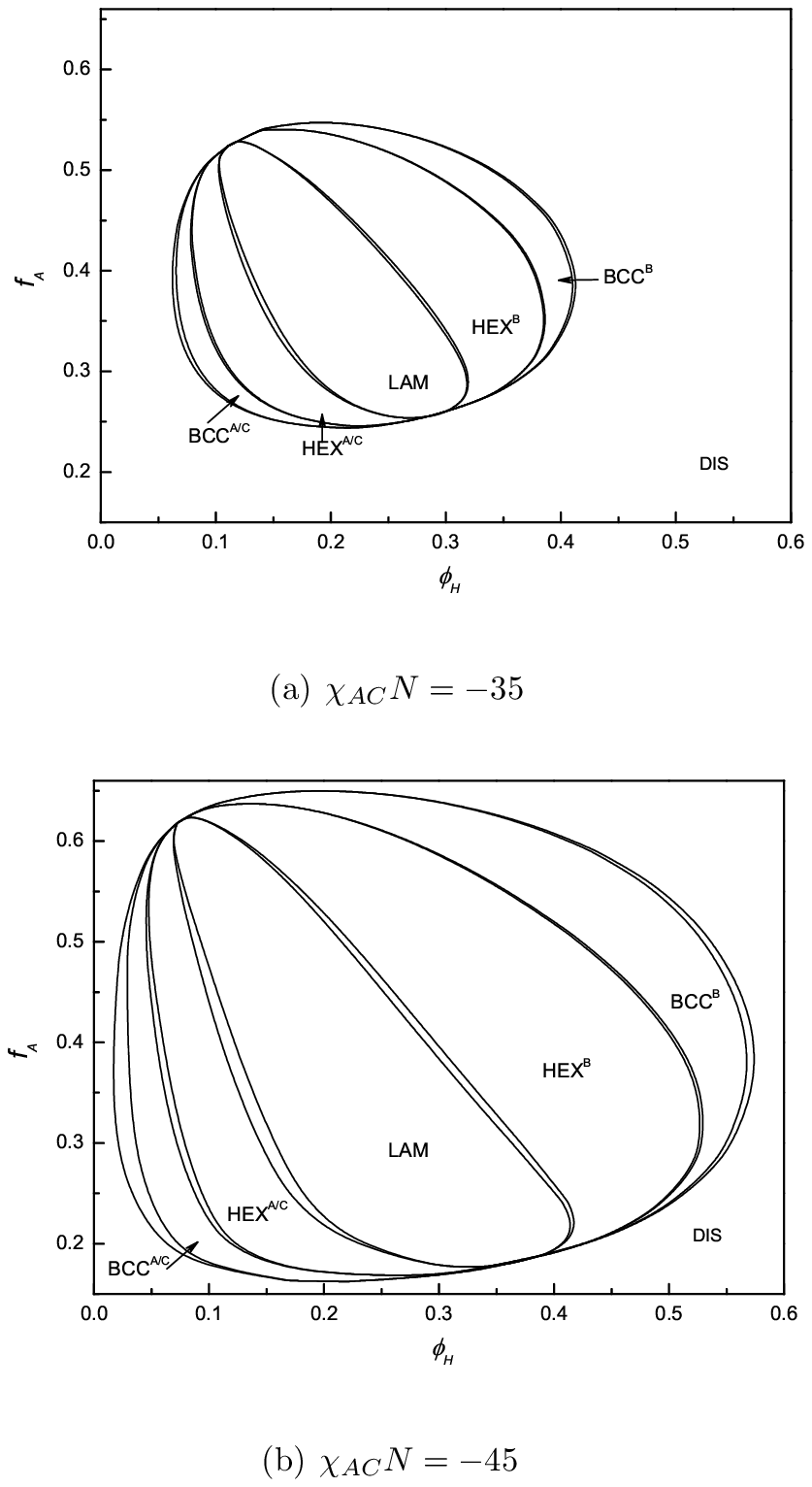}
  \caption{
    Phase diagrams of AB/C blends with parameters $\kappa=1.0$, 
    $\chi_{AB}N=2$, $\chi_{BC}N=0$, and (a) $\chi_{AC}N=-35$, (b)
    $\chi_{AC}N=-45$. 
  }
  \label{fig7}
\end{figure}

In order to explore the effect of the A/C interactions, phase diagrams
are constructed for different A/C interactions, either by changing the
molecular-weight of C homopolymers or by changing the value of the
Flory-Huggins parameter $\chi_{AC}$.  
Figure \ref{fig6}(a), \ref{fig5} and \ref{fig6}(b) show the
evolution of the phase diagrams with increasing $\kappa$
($\kappa=0.5$, $1.0$ and $1.5$).  
Alternatively, the phase diagrams in Figure \ref{fig7}(a),
\ref{fig5} and \ref{fig7}(b) demonstrate the progression as
$\chi_{AC}N$ decreases ($\chi_{AC}N=-35$, $-40$ and $-45$) at fixed
$\kappa$.  
It is obvious that the qualitative feature of the phase diagram stays
the same in these phase diagrams. 
The main difference is the size of the closed-loop ordered phase region. 
Increasing the interaction will increase the area of the parameter
space in which the blend is ordered.

Experimentally, several groups had studied blends with attractive 
interactions \cite{Chen2008, Tirumala2008}. 
In Ref. \cite{Chen2008}, Chen et al. studies the phase diagrams of 
poly(vinylphenol-\emph{b}-methyl methacrylate)/
poly(vinylpyrrolidone) (PVPh-\emph{b}-PMMA/PVP) blends, where all
three binary pairs, 
PVPh/PMMA, PMMA/PVP and PVP/PVPh, are miscible.  
Their setup closely resembles the blends studied in this work.  
They found a closed-loop microphase separation region 
surrounded by disordered phases.
The three classical phases were also observed inside the closed-loop.
The position of the closed-loop is similar to our theoretical
prediction. 
Furthermore, the transition sequence of ordered phases from their
experiments is consistent with our theoretical phase diagrams.

In a slightly different experiment setup \cite{Tirumala2008}, Tirumala
et al. studied a poly(oxyethylene-oxypropylene-oxyethylene)
triblock copolymer/poly(acrylic acid) blend (PEO-PPO-PEO/PAA),
where PAA interacts selectively to the end-block PEO. 
Upon increasing homopolymer concentration, they observed a phase
sequence of disordered phase to lamellae, then to cylinders consisted
of PPO, then back to disordered phase.
This transition sequence is again consistent with our theoretical
predictions.

\section{Summary}

In this paper, we have investigated the phase behavior of blends
composed of AB diblock copolymers and C homopolymers.  
Two theoretical methods have been employed to construct the phase
diagrams. 
By using the random phase approximation, the stability limits of the
homogeneous phase are obtained.  
The phase diagrams are characterized by the coexistence of macrophase
separation and microphase separation. 
When the interaction between A/C is repulsive, increasing $\chi_{AC}N$
will induce the transition from disordered phase to ordered phase, and
the property of the transition depends on the homopolymer concentration.
In general, blends with low homopolymer concentration tend to undergo
microphase separation transition, whereas macrophase separation occurs
at high homopolymer concentration. 
When the interaction parameter $\chi_{AC}N$ becomes attractive, the
macrophase separation disappears while the microphase separation can
still occur when the magnitude of $\chi_{AC}N$ is large.   
The microphase separation is caused by the strong attraction between the
homopolymers to one of the blocks of the diblock copolymers. 
The difference between the A/C and B/C interactions leads to the spatial
separation of B monomers from the A/C monomers, but the chemical
connections between A and B prevent the macrophase separation. 
The microphase separation manifests itself in the $\phi_H$-$f_A$
phase diagram in a closed-loop region.

The detailed morphologies inside the closed-loop are calculated
using self-consistent field theory.
The RPA results provide a guidance to explore the parameter space.
Three classical ordered phases (LAM, HEX and BCC) are included in the
current study. 
Along the closed-loop, the microphase separation transition is
a first-order phase transition from disordered phase to the microphase
with the symmetry of a body-centered cubic lattice except at two
critical points.
At these two points, the mean-field approach indicates that the blends
undergo continuous phase transition from disordered phase to lamellar
phase.  
When the homopolymer concentration is low, the phase sequence upon
increasing homopolymer concentration is from BCC to HEX, and from HEX
to LAM.  
At high homopolymer concentration, the phase sequence is reversed.

Another interesting observation of the phase diagram is the reverse
morphology where the majority component of the diblock forms the
dispersed domains of the ordered structure when small amount of 
homopolymers are added into the diblock melts. 
This phenomenon is due to the fact that the attractive interaction
of A/C is much stronger than that of B/C.   

The closed-loop microphase separation region has been observed in
experiment \cite{Chen2008}, and our theoretical prediction is
in qualitative agreement with the experimental result. 
Phase diagrams for blends with different homopolymer lengths and
Flory-Huggins parameters are also presented. 
It is observed that varying the interaction changes the size of the
closed-loop, while the general features of the phase diagram are
preserved.

\begin{acknowledgments}
This work was supported by the Natural Sciences and Engineering
Research Council (NSERC) of Canada. The computation was made possible
by the facilities of the Shared Hierarchical Academic Research
Computing Network (SHARCNET:www.sharcnet.ca).  
\end{acknowledgments}


\begin{thebibliography}{33}
\expandafter\ifx\csname natexlab\endcsname\relax\def\natexlab#1{#1}\fi
\expandafter\ifx\csname bibnamefont\endcsname\relax
  \def\bibnamefont#1{#1}\fi
\expandafter\ifx\csname bibfnamefont\endcsname\relax
  \def\bibfnamefont#1{#1}\fi
\expandafter\ifx\csname citenamefont\endcsname\relax
  \def\citenamefont#1{#1}\fi
\expandafter\ifx\csname url\endcsname\relax
  \def\url#1{\texttt{#1}}\fi
\expandafter\ifx\csname urlprefix\endcsname\relax\def\urlprefix{URL }\fi
\providecommand{\bibinfo}[2]{#2}
\providecommand{\eprint}[2][]{\url{#2}}

\bibitem[{\citenamefont{Paul and Newman}(1978)}]{Paul_Newman}
\bibinfo{editor}{\bibfnamefont{D.}~\bibnamefont{Paul}} \bibnamefont{and}
  \bibinfo{editor}{\bibfnamefont{S.}~\bibnamefont{Newman}}, eds.,
  \emph{\bibinfo{title}{Polymer Blends}} (\bibinfo{publisher}{Academic Press},
  \bibinfo{address}{New York}, \bibinfo{year}{1978}).

\bibitem[{\citenamefont{Utracki}(1989)}]{Utracki}
\bibinfo{author}{\bibfnamefont{L.}~\bibnamefont{Utracki}},
  \emph{\bibinfo{title}{Polymer Alloys and Blends: Thermodynamics and
  Rheology}} (\bibinfo{publisher}{Hanser Pub}, \bibinfo{address}{New York},
  \bibinfo{year}{1989}).

\bibitem[{\citenamefont{de~Gennes}(1979)}]{deGennes}
\bibinfo{author}{\bibfnamefont{P.-G.} \bibnamefont{de~Gennes}},
  \emph{\bibinfo{title}{Scaling Concepts in Polymer Physics}}
  (\bibinfo{publisher}{Cornell University Press}, \bibinfo{address}{Ithaca},
  \bibinfo{year}{1979}).

\bibitem[{\citenamefont{Leibler}(1980)}]{Leibler1980}
\bibinfo{author}{\bibfnamefont{L.}~\bibnamefont{Leibler}},
  \bibinfo{journal}{Macromolecules} \textbf{\bibinfo{volume}{13}},
  \bibinfo{pages}{1602} (\bibinfo{year}{1980}).

\bibitem[{\citenamefont{Hamley}(1998)}]{Hamley}
\bibinfo{author}{\bibfnamefont{I.}~\bibnamefont{Hamley}},
  \emph{\bibinfo{title}{The Physics of Block Copolymers}}
  (\bibinfo{publisher}{Oxford University Press}, \bibinfo{address}{Oxford},
  \bibinfo{year}{1998}).

\bibitem[{\citenamefont{Jiang and Xie}(1991)}]{Jiang1991}
\bibinfo{author}{\bibfnamefont{M.}~\bibnamefont{Jiang}} \bibnamefont{and}
  \bibinfo{author}{\bibfnamefont{H.}~\bibnamefont{Xie}},
  \bibinfo{journal}{Prog. Polym. Sci.} \textbf{\bibinfo{volume}{16}},
  \bibinfo{pages}{977} (\bibinfo{year}{1991}).

\bibitem[{\citenamefont{Jiang et~al.}(1995)\citenamefont{Jiang, Huang, and
  Xie}}]{Jiang1995}
\bibinfo{author}{\bibfnamefont{M.}~\bibnamefont{Jiang}},
  \bibinfo{author}{\bibfnamefont{T.}~\bibnamefont{Huang}}, \bibnamefont{and}
  \bibinfo{author}{\bibfnamefont{J.}~\bibnamefont{Xie}},
  \bibinfo{journal}{Macromol. Chem. Phys.} \textbf{\bibinfo{volume}{196}},
  \bibinfo{pages}{787} (\bibinfo{year}{1995}).

\bibitem[{\citenamefont{Lowenhaupt et~al.}(1994)\citenamefont{Lowenhaupt,
  Steurer, Hellmann, and Gallot}}]{Lowenhaupt1994}
\bibinfo{author}{\bibfnamefont{B.}~\bibnamefont{L\"owenhaupt}},
  \bibinfo{author}{\bibfnamefont{A.}~\bibnamefont{Steurer}},
  \bibinfo{author}{\bibfnamefont{G.}~\bibnamefont{Hellmann}}, \bibnamefont{and}
  \bibinfo{author}{\bibfnamefont{Y.}~\bibnamefont{Gallot}},
  \bibinfo{journal}{Macromolecules} \textbf{\bibinfo{volume}{27}},
  \bibinfo{pages}{908} (\bibinfo{year}{1994}).

\bibitem[{\citenamefont{Zhao et~al.}(1997)\citenamefont{Zhao, Pearce, and
  Kwei}}]{Zhao1997}
\bibinfo{author}{\bibfnamefont{J.}~\bibnamefont{Zhao}},
  \bibinfo{author}{\bibfnamefont{E.}~\bibnamefont{Pearce}}, \bibnamefont{and}
  \bibinfo{author}{\bibfnamefont{T.}~\bibnamefont{Kwei}},
  \bibinfo{journal}{Macromolecules} \textbf{\bibinfo{volume}{30}},
  \bibinfo{pages}{7119} (\bibinfo{year}{1997}).

\bibitem[{\citenamefont{Han et~al.}(2000)\citenamefont{Han, Pearce, and
  Kwei}}]{Han2000}
\bibinfo{author}{\bibfnamefont{Y.}~\bibnamefont{Han}},
  \bibinfo{author}{\bibfnamefont{E.}~\bibnamefont{Pearce}}, \bibnamefont{and}
  \bibinfo{author}{\bibfnamefont{T.}~\bibnamefont{Kwei}},
  \bibinfo{journal}{Macromolecules} \textbf{\bibinfo{volume}{33}},
  \bibinfo{pages}{1321} (\bibinfo{year}{2000}).

\bibitem[{\citenamefont{Semenov}(1993)}]{Semenov1993}
\bibinfo{author}{\bibfnamefont{A.}~\bibnamefont{Semenov}},
  \bibinfo{journal}{Macromolecules} \textbf{\bibinfo{volume}{26}},
  \bibinfo{pages}{2273} (\bibinfo{year}{1993}).

\bibitem[{\citenamefont{Matsen}(1995{\natexlab{a}})}]{Matsen1995}
\bibinfo{author}{\bibfnamefont{M.}~\bibnamefont{Matsen}},
  \bibinfo{journal}{Phys. Rev. Lett.} \textbf{\bibinfo{volume}{74}},
  \bibinfo{pages}{4225} (\bibinfo{year}{1995}{\natexlab{a}}).

\bibitem[{\citenamefont{Matsen}(1995{\natexlab{b}})}]{Matsen1995a}
\bibinfo{author}{\bibfnamefont{M.}~\bibnamefont{Matsen}},
  \bibinfo{journal}{Macromolecules} \textbf{\bibinfo{volume}{28}},
  \bibinfo{pages}{5765} (\bibinfo{year}{1995}{\natexlab{b}}).

\bibitem[{\citenamefont{Janert and Schick}(1998)}]{Janert1998}
\bibinfo{author}{\bibfnamefont{P.}~\bibnamefont{Janert}} \bibnamefont{and}
  \bibinfo{author}{\bibfnamefont{M.}~\bibnamefont{Schick}},
  \bibinfo{journal}{Macromolecules} \textbf{\bibinfo{volume}{31}},
  \bibinfo{pages}{1109} (\bibinfo{year}{1998}).

\bibitem[{\citenamefont{Hashimoto et~al.}(1990)\citenamefont{Hashimoto, Tanaka,
  and Hasegawa}}]{Hashimoto1990}
\bibinfo{author}{\bibfnamefont{T.}~\bibnamefont{Hashimoto}},
  \bibinfo{author}{\bibfnamefont{H.}~\bibnamefont{Tanaka}}, \bibnamefont{and}
  \bibinfo{author}{\bibfnamefont{H.}~\bibnamefont{Hasegawa}},
  \bibinfo{journal}{Macromolecules} \textbf{\bibinfo{volume}{23}},
  \bibinfo{pages}{4378} (\bibinfo{year}{1990}).

\bibitem[{\citenamefont{Tanaka et~al.}(1991)\citenamefont{Tanaka, Hasegawa, and
  Hashimoto}}]{Tanaka1991}
\bibinfo{author}{\bibfnamefont{H.}~\bibnamefont{Tanaka}},
  \bibinfo{author}{\bibfnamefont{H.}~\bibnamefont{Hasegawa}}, \bibnamefont{and}
  \bibinfo{author}{\bibfnamefont{T.}~\bibnamefont{Hashimoto}},
  \bibinfo{journal}{Macromolecules} \textbf{\bibinfo{volume}{24}},
  \bibinfo{pages}{240} (\bibinfo{year}{1991}).

\bibitem[{\citenamefont{Winey et~al.}(1991)\citenamefont{Winey, Thomas, and
  Fetters}}]{Winey1991}
\bibinfo{author}{\bibfnamefont{K.}~\bibnamefont{Winey}},
  \bibinfo{author}{\bibfnamefont{E.}~\bibnamefont{Thomas}}, \bibnamefont{and}
  \bibinfo{author}{\bibfnamefont{L.}~\bibnamefont{Fetters}},
  \bibinfo{journal}{Macromolecules} \textbf{\bibinfo{volume}{24}},
  \bibinfo{pages}{6182} (\bibinfo{year}{1991}).

\bibitem[{\citenamefont{Zeman and Patterson}(1972)}]{Zeman1972}
\bibinfo{author}{\bibfnamefont{L.}~\bibnamefont{Zeman}} \bibnamefont{and}
  \bibinfo{author}{\bibfnamefont{D.}~\bibnamefont{Patterson}},
  \bibinfo{journal}{Macromolecules} \textbf{\bibinfo{volume}{5}},
  \bibinfo{pages}{513} (\bibinfo{year}{1972}).

\bibitem[{\citenamefont{Jo et~al.}(1991)\citenamefont{Jo, Kwon, and
  Kwon}}]{Jo1991}
\bibinfo{author}{\bibfnamefont{W.}~\bibnamefont{Jo}},
  \bibinfo{author}{\bibfnamefont{Y.}~\bibnamefont{Kwon}}, \bibnamefont{and}
  \bibinfo{author}{\bibfnamefont{I.}~\bibnamefont{Kwon}},
  \bibinfo{journal}{Macromolecules} \textbf{\bibinfo{volume}{24}},
  \bibinfo{pages}{4708} (\bibinfo{year}{1991}).

\bibitem[{\citenamefont{Manestrel et~al.}(1992)\citenamefont{Manestrel,
  Bhagwagar, Painter, Coleman, and Graf}}]{Manestrel1992}
\bibinfo{author}{\bibfnamefont{C.}~\bibnamefont{Manestrel}},
  \bibinfo{author}{\bibfnamefont{D.}~\bibnamefont{Bhagwagar}},
  \bibinfo{author}{\bibfnamefont{P.}~\bibnamefont{Painter}},
  \bibinfo{author}{\bibfnamefont{M.}~\bibnamefont{Coleman}}, \bibnamefont{and}
  \bibinfo{author}{\bibfnamefont{J.}~\bibnamefont{Graf}},
  \bibinfo{journal}{Macromolecules} \textbf{\bibinfo{volume}{25}},
  \bibinfo{pages}{1701} (\bibinfo{year}{1992}).

\bibitem[{\citenamefont{Kuo et~al.}(2002)\citenamefont{Kuo, Lin, and
  Chang}}]{Kuo2002}
\bibinfo{author}{\bibfnamefont{S.-W.} \bibnamefont{Kuo}},
  \bibinfo{author}{\bibfnamefont{C.-L.} \bibnamefont{Lin}}, \bibnamefont{and}
  \bibinfo{author}{\bibfnamefont{F.-C.} \bibnamefont{Chang}},
  \bibinfo{journal}{Macromolecules} \textbf{\bibinfo{volume}{35}},
  \bibinfo{pages}{278} (\bibinfo{year}{2002}).

\bibitem[{\citenamefont{Chen et~al.}(2008)\citenamefont{Chen, Kuo, Jeng, and
  Chang}}]{Chen2008}
\bibinfo{author}{\bibfnamefont{W.-C.} \bibnamefont{Chen}},
  \bibinfo{author}{\bibfnamefont{S.-W.} \bibnamefont{Kuo}},
  \bibinfo{author}{\bibfnamefont{U.-S.} \bibnamefont{Jeng}}, \bibnamefont{and}
  \bibinfo{author}{\bibfnamefont{F.-C.} \bibnamefont{Chang}},
  \bibinfo{journal}{Macromolecules} \textbf{\bibinfo{volume}{41}},
  \bibinfo{pages}{1401} (\bibinfo{year}{2008}).

\bibitem[{\citenamefont{Mori et~al.}(1987)\citenamefont{Mori, Tanaka, and
  Hashimoto}}]{Mori1987}
\bibinfo{author}{\bibfnamefont{K.}~\bibnamefont{Mori}},
  \bibinfo{author}{\bibfnamefont{H.}~\bibnamefont{Tanaka}}, \bibnamefont{and}
  \bibinfo{author}{\bibfnamefont{T.}~\bibnamefont{Hashimoto}},
  \bibinfo{journal}{Macromolecules} \textbf{\bibinfo{volume}{20}},
  \bibinfo{pages}{381} (\bibinfo{year}{1987}).

\bibitem[{\citenamefont{Ijichi and Hashimoto}(1988)}]{Ijichi1988}
\bibinfo{author}{\bibfnamefont{Y.}~\bibnamefont{Ijichi}} \bibnamefont{and}
  \bibinfo{author}{\bibfnamefont{T.}~\bibnamefont{Hashimoto}},
  \bibinfo{journal}{Polym. Commun.} \textbf{\bibinfo{volume}{29}},
  \bibinfo{pages}{135} (\bibinfo{year}{1988}).

\bibitem[{\citenamefont{Tanaka and Hashimoto}(1988)}]{Tanaka1988}
\bibinfo{author}{\bibfnamefont{H.}~\bibnamefont{Tanaka}} \bibnamefont{and}
  \bibinfo{author}{\bibfnamefont{T.}~\bibnamefont{Hashimoto}},
  \bibinfo{journal}{Polym. Commun.} \textbf{\bibinfo{volume}{29}},
  \bibinfo{pages}{212} (\bibinfo{year}{1988}).

\bibitem[{\citenamefont{Matsen and Schick}(1994)}]{Matsen1994}
\bibinfo{author}{\bibfnamefont{M.}~\bibnamefont{Matsen}} \bibnamefont{and}
  \bibinfo{author}{\bibfnamefont{M.}~\bibnamefont{Schick}},
  \bibinfo{journal}{Phys. Rev. Lett.} \textbf{\bibinfo{volume}{72}},
  \bibinfo{pages}{2660} (\bibinfo{year}{1994}).

\bibitem[{\citenamefont{Shi}(2004)}]{Shi2004_chapter}
\bibinfo{author}{\bibfnamefont{A.-C.} \bibnamefont{Shi}}, in
  \emph{\bibinfo{booktitle}{Developments in Block Copolymer Science and
  Technology}}, edited by
  \bibinfo{editor}{\bibfnamefont{I.}~\bibnamefont{Hamley}}
  (\bibinfo{publisher}{John Wiley \& Sons}, \bibinfo{address}{New York},
  \bibinfo{year}{2004}), chap.~\bibinfo{chapter}{8}.

\bibitem[{\citenamefont{Fredrickson}(2006)}]{Fredrickson}
\bibinfo{author}{\bibfnamefont{G.~H.} \bibnamefont{Fredrickson}},
  \emph{\bibinfo{title}{The Equilibrium Theory of Inhomogeneous Polymers}}
  (\bibinfo{publisher}{Clarendon Press}, \bibinfo{address}{Oxford},
  \bibinfo{year}{2006}).

\bibitem[{\citenamefont{Laradji et~al.}(1997)\citenamefont{Laradji, Shi, Desai,
  and Noolandi}}]{Laradji1997}
\bibinfo{author}{\bibfnamefont{M.}~\bibnamefont{Laradji}},
  \bibinfo{author}{\bibfnamefont{A.-C.} \bibnamefont{Shi}},
  \bibinfo{author}{\bibfnamefont{R.}~\bibnamefont{Desai}}, \bibnamefont{and}
  \bibinfo{author}{\bibfnamefont{J.}~\bibnamefont{Noolandi}},
  \bibinfo{journal}{Phys. Rev. Lett.} \textbf{\bibinfo{volume}{78}},
  \bibinfo{pages}{2577} (\bibinfo{year}{1997}).

\bibitem[{\citenamefont{Wang et~al.}(2009)\citenamefont{Wang, Li, Luo, Li, Shi,
  and Zhu}}]{Wang2009}
\bibinfo{author}{\bibfnamefont{R.}~\bibnamefont{Wang}},
  \bibinfo{author}{\bibfnamefont{W.}~\bibnamefont{Li}},
  \bibinfo{author}{\bibfnamefont{Y.}~\bibnamefont{Luo}},
  \bibinfo{author}{\bibfnamefont{B.-G.} \bibnamefont{Li}},
  \bibinfo{author}{\bibfnamefont{A.-C.} \bibnamefont{Shi}}, \bibnamefont{and}
  \bibinfo{author}{\bibfnamefont{S.}~\bibnamefont{Zhu}},
  \bibinfo{journal}{Macromolecules} \textbf{\bibinfo{volume}{42}},
  \bibinfo{pages}{2275} (\bibinfo{year}{2009}).

\bibitem[{\citenamefont{Fredrickson and Helfand}(1987)}]{Fredrickson1987}
\bibinfo{author}{\bibfnamefont{G.~H.} \bibnamefont{Fredrickson}}
  \bibnamefont{and} \bibinfo{author}{\bibfnamefont{E.}~\bibnamefont{Helfand}},
  \bibinfo{journal}{J. Chem. Phys.} \textbf{\bibinfo{volume}{87}},
  \bibinfo{pages}{697} (\bibinfo{year}{1987}).

\bibitem[{\citenamefont{Kielhorn and Muthukumar}(1997)}]{Kielhorn1997}
\bibinfo{author}{\bibfnamefont{L.}~\bibnamefont{Kielhorn}} \bibnamefont{and}
  \bibinfo{author}{\bibfnamefont{M.}~\bibnamefont{Muthukumar}},
  \bibinfo{journal}{J. Chem. Phys.} \textbf{\bibinfo{volume}{107}},
  \bibinfo{pages}{5588} (\bibinfo{year}{1997}).

\bibitem[{\citenamefont{Tirumala et~al.}(2008)\citenamefont{Tirumala, Daga,
  Bosse, Romang, Ilavsky, Lin, and Watkins}}]{Tirumala2008}
\bibinfo{author}{\bibfnamefont{V.}~\bibnamefont{Tirumala}},
  \bibinfo{author}{\bibfnamefont{V.}~\bibnamefont{Daga}},
  \bibinfo{author}{\bibfnamefont{A.}~\bibnamefont{Bosse}},
  \bibinfo{author}{\bibfnamefont{A.}~\bibnamefont{Romang}},
  \bibinfo{author}{\bibfnamefont{J.}~\bibnamefont{Ilavsky}},
  \bibinfo{author}{\bibfnamefont{E.}~\bibnamefont{Lin}}, \bibnamefont{and}
  \bibinfo{author}{\bibfnamefont{J.}~\bibnamefont{Watkins}},
  \bibinfo{journal}{Macromolecules} \textbf{\bibinfo{volume}{41}},
  \bibinfo{pages}{7978} (\bibinfo{year}{2008}).

\end{thebibliography}

\begin{center}
***
\end{center}

\footnotesize{--- \emph{Copyright 2009 American Institute of
    Physics. This article may be downloaded for personal use only. Any
    other use requires prior permission of the author and the American
    Institute of Physics.} 
}

\footnotesize{--- \emph{The following article appeared in
    J. Chem. Phys. \textbf{130}, 234904 (2009) and may be found at
    http://link.aip.org/link/?JCPSA6/130/234904/1.}
}

\end{document}